\documentclass[conference]{IEEEtran}
\ifCLASSINFOpdf
  \usepackage[pdftex]{graphicx}
\else
\fi
\usepackage{url}

\begin{document}
%
\title{Adaptive Automation: Leveraging Machine Learning to Support Uninterrupted Automated Testing of Software Applications}

\author{\IEEEauthorblockN{Rajesh Mathur, Scott Miles, Miao Du}
\IEEEauthorblockA{Awesomation Testing\\
Melbourne, Australia\\
\url{http://www.awesomationtesting.com}\\
Email: \{rmathur,smiles,mdu\}@awesomationtesting.com}
}


%


\maketitle

\begin{abstract}
Checking software application suitability using automated software tools has become a vital element for most organisations irrespective of whether they produce in-house software or simply customise off-the-shelf software applications for internal use. As software solutions become ever more complex, the industry becomes increasingly  dependent on software automation tools, yet the brittle nature of the available software automation tools limits their effectiveness. Companies invest significantly in obtaining and implementing automation software but most of the tools fail to deliver when the cost of maintaining an effective automation test suite exceeds the cost and time that would have otherwise been spent on manual testing. A failing in the current generation of software automation tools is they do not adapt to unexpected modifications and obstructions without frequent (and time expensive) manual interference. Such issues are commonly acknowledged amongst industry practitioners, yet none of the current generation of tools have leveraged the advances in machine learning and artificial intelligence to address these problems.

This paper proposes a framework solution that utilises machine learning concepts, namely fuzzy matching and error recovery. The suggested solution applies adaptive techniques to recover from unexpected obstructions that would otherwise have prevented the script from proceeding. Recovery details are presented to the user in a report which can be analysed to determine if the recovery procedure was acceptable and the framework will adapt future runs based on the decisions of the user. Using this framework, a practitioner can run the automated suits without human intervention while minimising the risk of schedule delays. 
\end{abstract}



%

\section{Introduction}

In the last decade, the Information Technology industry has grown multifold. What we could not imagine pre-internet era is now available to the masses. Laptop and Desktop computers with gigabytes of memory and terabytes of hard disk space are available to school kids. Smartphones that have capabilities of early age supercomputers are adorned by common people. Wireless network has become a necessity and is being considered as one of the core components of Maslow’ Need Hierarchy theory. Software applications that were unimaginable at the turn of the century are now available and are becoming increasingly complex. 

Software testing that was once considered a repetitive and less-intelligent activity is proving to be one of the most important activities during the Software Development Life Cycle. The software testing services industry is forecast to grow at greater than 11\% annually \cite{technavio:2014} and a recent industry report suggests that the industry will be a \$US46 billion industry within three years \cite{MicroFocus:2011}.

The increasing complexity of the software application combined with the increasing demand for shorter release cycles has led to the expectation of more testing within an ever shortening time frame. In order to accomplish this goal organisations often turn to support from computers that can check their own applications. The use of software automation tools to test an application has now become a norm in the software industry. While testing by humans is considered a complex cognitive activity, the effectiveness of it can be greatly enhanced when coupled with automation to perform the tasks that are repetitive in nature and require frequent regression. When used for the task of checking, automation tools can provide a fast and consistent approach to support the conformance of a product to business requirements. 

One of the biggest challenges that the software industry faces while using commercial automation tools is that most of these tools are unintelligent and require significant maintenance. Almost all of these tools focus solely on conforming to information that has been inputted by a human user and are forced to abort a test when an obstruction occurs. Unfortunately human users are all too susceptible to error and will commonly input the wrong thing (expecting something that should not be expected), forget to input something (not updating an existing script with an intended change) or will not account for something (users on a different browser version will get an additional popup dialog when accessing certain webpages). The software industry has not grown in developing intelligent tools or equally intelligent users of these tools. Many companies depend on record and replay features and trust that the testing will be done completely with sufficient coverage \cite{kaner:1997}.

\section{Test Automation Challenges}

The promised benefits of test automation are undermined by serious limitations in the current tools and methodology.  We describe four major challenges with the current tools, particularly with respect to automated user interface testing.

\subsection*{Challenge 1: Test Case Generation is Manual and Labour Intensive}

There are many different approaches used for generation of test scripts within the current generation of automation tools. The majority of tools rely on a tester to script what actions should occur and what checks to perform, or to record actions to be replayed in the exact sequence on demand. The tools make no attempt at understanding the intended function of a system and generate an automatic suite of possible test scripts. While a tool may not be able to determine all of the expected behavior of a system, there are many aspects that are common to all systems or can be learned based on previously scripted tests.

\subsection*{Challenge 2: Test Scripts are Not Validated until Runtime}

Testing tools commonly use an \emph{object repository} to map a user friendly object name with a list of object parameters used to identify the object. When a script is written within a test tool, the majority of tools offer no live validation that items referred to in the test script exist in the repository. For example if the command to click the object \emph{OK Button} is scripted, there is no validation of whether the \emph{OK Button} exists in the repository until the script is executed. Likewise, if the test author enters the command to ``Enter Text'' on the button, it again may not be detected as an invalid command until the script is executed.

Some tools attempt to address this limitation by verifying the objects used in a script when it is created and modified, however if the item is later removed or modified from the object repository there may not be any indication if a script has been broken.

\subsection*{Challenge 3: Test Scripts are Brittle}

When an automation script fails to locate an item on the current screen the majority of frameworks will do one of two things (depending on the framework and settings used):
\begin{itemize}
\item Skip remaining steps and fail the whole test.
\item Fail the single step and continue to next step which will most likely fail as a result of the missed step.
\end{itemize}

Often these failures occur when an object has a code change which affects the parameters being used to identify the object. Examples of such changes include: changing of the name parameter, changing the type of an object, or changing the parent container. Failing a test because of such a change such is generally not the desired outcome as only a GUI element has changed, there is no effect on the functionality of the system. If the element has changed the tool will no longer be able to be identify the element and therefore the test will not perform the intended action, potentially causing a cascade of failures.  Maintenance of these changes is costly and involves a significant amount of investigation, updating and repetition of sequential tests in order to verify the corrected script.

Another cause of failure can be slight cosmetic changes, such as: changing a button to rounded corners, changing the text on a hyperlink, changing the location of a button by 1 pixel, or changing the background in a screenshot comparison. While these changes appear trivial to the application developer and end user, these changes impact the tester, as they can break the automated test case.

A third cause of failure is an unexpected occurrence. These often occur as popup windows, warning dialogs, or additional steps. Some examples of these are: the webpage security certificate has expired and must be accepted, payment was made recently with the same credit card or, accepting changed terms and conditions before proceeding. The combinations of unexpected occurrences that could occur in different locations of a script are innumerable making these scenarios difficult to predict and costly to accomodate.

The last frequent cause of automation scripts breaking is failure of an action to load completely within a timely fashion. Often a database query will hang without returning a result or a webpage will not be served correctly on the first attempt. While it may be significant depending on the circumstances and the testing being performed this is often due to testing on restricted infrastructure within a test environment. The functionality itself has not been affected and the test would be able to continue if the framework could recover from such an event.

All of these types of failures should be brought to the tester’s attention to determine if the change is acceptable, but should not cause a test to stop or subsequent steps to fail where an alternative is available.

\subsection*{Challenge 4: Maintaining a Test Suite is Expensive}
Automation is most commonly applied as a regression check for features that have reached a mature state in their life cycle. When changes occur to a mature feature, they are usually minor, such as adding a new field to a drop down box, changing the text in a paragraph, adding a link to a new page on a website.

The functionality of these changes can usually be assumed based on the existing test scripts and the context of the change, yet can require a lot of effort on the testers behalf to update every script that touches that page. For example, if an online store adds a new size option to their ‘Add to Cart’ functionality each test script that touches that function would need to include checks for this option.

It is also possible for a change to be implemented and tested manually with the parties involved forgetting to ensure that the automation framework has been updated. This will often cause a series of failures that were expected and then a rushed updating of the existing test scripts.
Tests must be expanded by the tester when a new option or field is included.

\begin{figure*}[htb]
\label{fig:framework}
\centering
\includegraphics[width=5in]{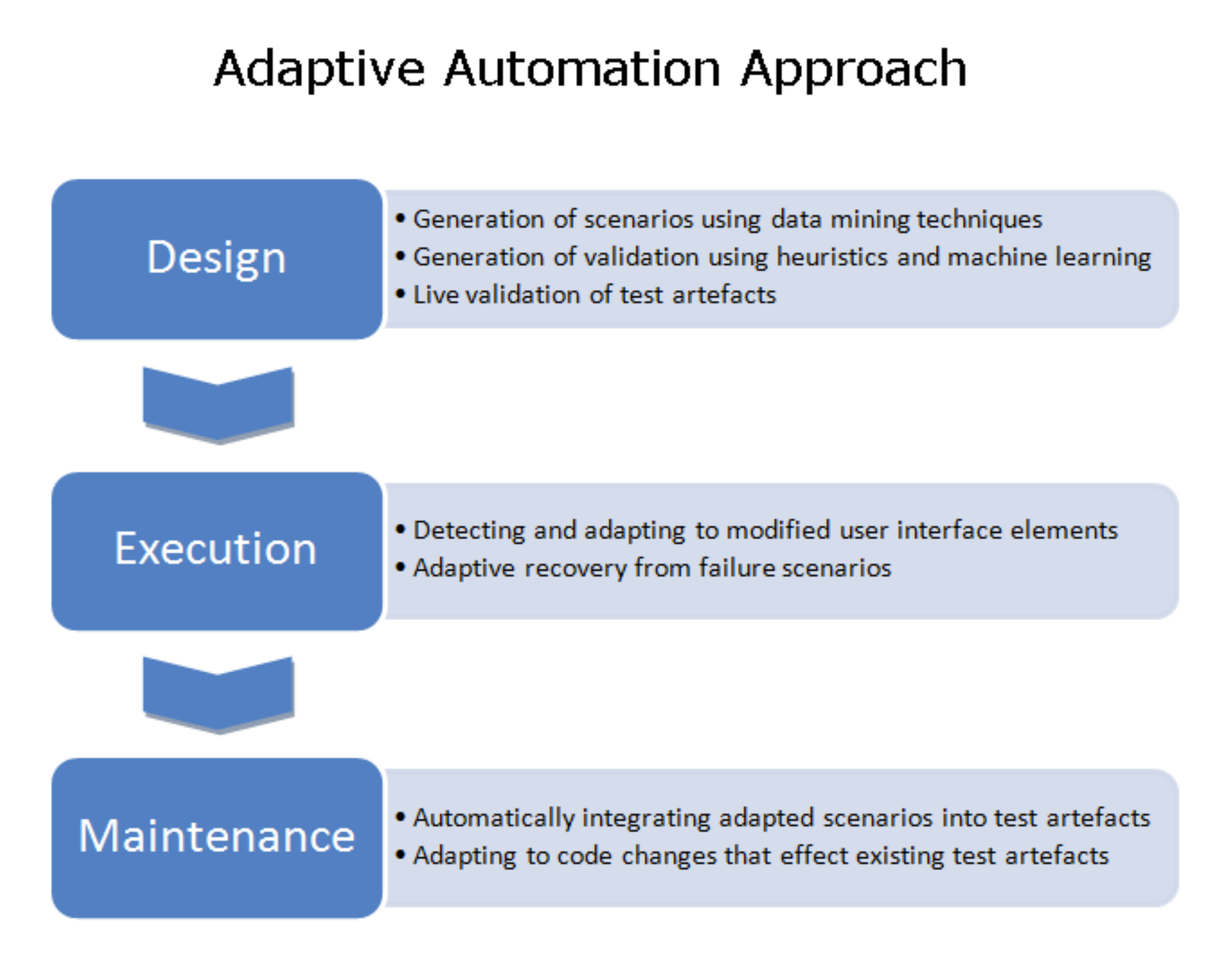}
\caption{The Adaptive Automation Approach}
\end{figure*}

\section{Adaptive Test Automation Roadmap}

To address these shortcomings in the current generation of automation tools, we propose that test automation frameworks should leverage self adaptive technologies and machine learning methods. A self adaptive automated test framework is needed to support the design, execution and maintenance of test cases as shown in Figure~\ref{fig:framework}. The remainder of this section proposes some the adaptive support mechanisms that could be given across various stages of the test automation lifecycle.

\subsection*{Goal 1: Test Generation - automatic test generation from screens and user behaviour}

Test scripts should be generated automatically where possible.  There are two general strategies which may be employed for automatic user interface test generation:

\begin{itemize}
\item
Automatically exploring the screens or pages and generating tests which exercise all of the discovered fields.  Heuristics may be used guide the values used in the test cases.  One could employ a database of common field names and associated formats.  For example if a field contains ``email'' in its name then check that it only accepts email formats. If a field is called ``age'', then an appropriate check would be that the field does not accept negative values.
\item
Data mining the user interaction logs to create test cases.  The values used in the test cases and the paths tested would be drawn directly from the user data.  This approach has the benefit of focusing on the use case scenarios most important to the users rather than theoretical execution paths which might rarely be exercised in practice.
\end{itemize}

\subsection*{Goal 2: Live Validation - dynamic checking of test scripts at authoring time}
Test case authoring tools should provide the same level of dynamic checking and assistance as is given by other software interactive development environments (such as Eclipse.)  Specifically:
\begin{itemize}
\item
If the script makes reference to a deleted or misspelled object or be missing a parameter, this should be detected at authoring time and flagged to the script author.  This would prevent many runtime test failures.
\item
Script authoring tools should know what actions are possible, and what parameters are valid for each action. These would be provided by dropdown lists and predictive text.
\item
If an object is changed or removed in a way that would cause an existing script to fail, a notification should occur immediately.
\end{itemize}

\subsection*{Goal 3: Adaptive Recovery - self adaptive recovery to test script interruptions}

To address the challenge of test script brittleness, we propose the following collection of techniques:
\begin{itemize}
\item
\emph{Fuzzy matching}: when an object required by the test script does not exist, rather than failing because there is no exact match, the script should adaptively recover by applying a fuzzy match to find the closest item to the expected object.  The recovery technique would try to find another object which has the closest match to the attributes of the missing object.  To define closest, many alternative similarity measures could be used.  If the attributes are numerical, the Euclidean distance or cosine distance might be used.  For string attributes, the edit distance could be used.  To calculate the overall similarity of two objects, the weighted sum of their attribute similarity could be applied, or a more sophisticated method, such as the Analytical Hierarchical Process.

As a simple concrete example, consider a script which expects a listbox called ``CitiesDropDown''.  When the expected object does not exist, the script might then try to use the another listbox called “CitiesList” instead.

\item
\emph{Check for unexpected popups}: when script is stalled, the proposed framework would automatically detect whether there are any unexpected popup windows and attempt to resolve the popup in the most suitable method. A popup definition would be frequently updated with known popups and recommended action to resolve the dialog. As tests are automated and client specific popups are encountered these would be added to the definition and machine learning would be applied to determine the resolution.

\item
\emph{Perform recovery techniques}: when the expected objects do not appear on the current page, the framework should determine possible recovery techniques and attempt these in order to continue the test. Some basic examples of recovery techniques would include:
(a) if the current page appears to be a login page, attempt to log in with previous details as user session may have expired;
(b) If a next or previous button is available, attempt to explore the adjacent pages for the object as author may have forgotten this step;
(c) if page has not loaded correctly, refresh the page and try again; and
(d) other custom actions determined by user or learnt from test scripts.

When the test script adaptively recovers from an unexpected condition, this should be reported.  The tester may then assess whether the adaptation was valid.  The tester may also wish to approve the adaptation being permanently added to the test script, or manually update the script for the changed conditions.

\end{itemize}

\subsection*{Goal 4: Automatic Maintenance - automatic updates of test scripts for application changes}
The final goal of adaptive automated testing is to link testing scripts directly to changes made to the application under test.

When the application under test version control system (such as Git) is available, changes to the user interface could be automatically detected.  The types of changes which affect the test script include: changes to the name of a referenced object, a referenced object is deleted or a new user interface object is added or otherwise changed.  The test owner shown be alerted to any such detected changes.  The ultimate goal of adaptive automated testing would be for such changes to automatically update the associated test scripts, with the test owner only asked to confirm the changes.  Automated updates to the script would include updating changed object names, removing code referencing deleted elements and expanding the test for the new elements.

An alternative approach, where the application under test version control system is not available is to automatically analyse the user interface for new versions of the application binary or SaaS delivered application.  The same kinds of information can be inferred by discovering the elements in the user interface screens and pages, and comparing it to the previous version.  The actionable steps, such as informing the test owner or automatically updating the test scripts would be the same as with a source control system.

When a new element is detected in a page that belongs to a specified script the framework could take the information it already has about this field to modify existing test scripts. For example If an online store adds a new size option to their ``Add to Cart'' feature, the tool already knows that when Small, Medium or Large is selected that the corresponding value is added to the cart, it would be a simple matter for it to recognise the new ``Extra Large'' option and apply the same logic.

\begin{figure*}[tb!]
\label{fig:proto}
\centering
\includegraphics[width=5in]{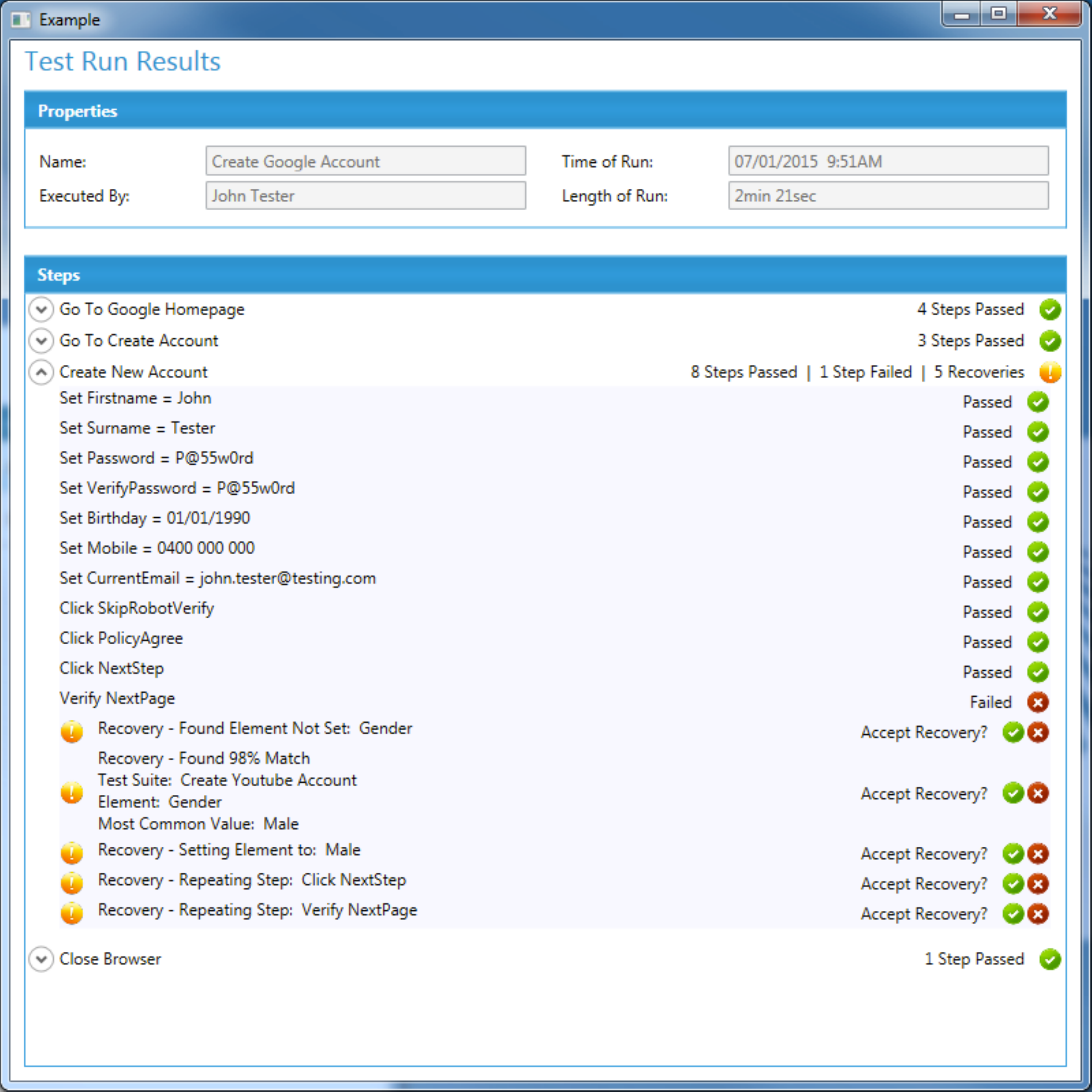}
\caption{A screenshot of the adaptive automation framework prototype showing the results of a test run with adaptive error recovery.}
\end{figure*}

\section{Prototype}

We have a proof-of-concept prototype which implements some of our goals of adaptive test recovery for automated user interface testing. The prototype addresses three common issues: missing user interface elements, disruptions to the user interface flow caused by pop-up dialogs and recovery of tests using pre-defined recovery techniques.

As an example of how the prototype works, suppose XYZ Corp is a fictional organisation which uses Google Apps.  XYZ Corp has a test script to validate the onboarding of their new employees.  Now imagine a scenario where the developers at Google have added an additional field to the Google account creation process where a user must enter their gender. The test script was never updated to include this additional step, and if this test was to fail in the current generation of Automation Tools the result would cascade to a large set of test cases that expect to use the newly created account. Figure~\ref{fig:proto} shows what would happen in the Adaptive Automation proposed solution.

The recovery feature of this framework has detected that the test case failure was most likely caused due to an element that was not set. It then proceeds to combine a set of logical heuristics along with what it can learn from similar elements within similar test cases and determines the best process to recover from this error. The remaining tests within the test suite can then proceed as scheduled and the tester can review the results to determine if the recovery process was acceptable. Future recovery attempts would be adjusted to compensate for the users’ expectations based on their decision to accept or reject the recovery attempt.

\section{Related Work}

HP Unified Functional Testing (formerly HP QTP) is possibly the most well known tool available in today's market. HP UFT offers a rich set of features in a polished graphical user interface including: test flow viewer, keyword driven tests, scripted tests, capture replay and many more.

Selenium is a well known open source test automation tool which focuses on testing web applications. It allows you to develop tests through recording tools or through manual scripting, and to execute those tests on multiple browsers across various platforms.

Cucumber is another well known open source framework offering a behavior-driven development approach to test automation. It is a Ruby language framework that is both powerful and simple at the same time allowing tests to be described through plain language scenarios.

\section{Discussion}

Automated testing is an important enabler for the rapid transfer of a software update from Development to Operations. However the current state of automated testing is limited because the time investment required at the test authoring and maintenance stages often outweighs the time savings yielded from automatic test execution. Automatic test scripts are brittle to small changes in the application under test or the test environment. This causes automatic testing to be unreliable and high maintenance.

We have proposed a framework for adaptive automated testing. Our framework supports automated testing at three stages of the test automation lifecycle: test design, test execution and test maintenance. Our framework automatically generates tests directly from the application under test reducing the cost of implementation. Test case authoring is supported with live validation of referenced objects in the test scripts. Test execution is made robust by self adaptive recovery, using a combination of techniques to recover from missing elements or unexpected events. Test maintenance is automated or semi-automated by directly linking changes to the application to updates in the test script.

We have developed a proof of concept prototype which demonstrates self adaptive recovery.  Our prototype applies fuzzy matching to find the closest match in the event of missing elements. It also automatically responds to unexpected events such as popup warning windows and applies recovery techniques to failing tests.

The focus of this paper has been on automated user interface driven testing. Our techniques could also be adapted to work with other forms of testing, such as API testing, performance testing and accessibility testing.


\section*{Acknowledgment}

The authors thank Professor John Grundy for the very informative discussions and suggested directions.

\IEEEtriggeratref{8}
\IEEEtriggercmd{\enlargethispage{-5in}}


\bibliographystyle{IEEEtran}
\bibliography{adaptive_auto}

\end{document}